\documentclass[sigconf]{acmart}
\usepackage{url}
\usepackage[normalem]{ulem}
\usepackage{amsmath,amssymb,amsfonts}
\usepackage{graphicx}
\usepackage{listings}
\usepackage{color}
\usepackage{courier}
\usepackage{float}
\usepackage{makecell}
\usepackage{algorithm} 
\usepackage[noend]{algpseudocode}
\usepackage[capitalize]{cleveref}

\acmYear{2019}

\begin{document}
\title{GraphCage: Cache Aware Graph Processing on GPUs}
\author{Xuhao Chen}
\affiliation{University of Texas at Austin}
\email{cxh@utexas.edu}
\begin{abstract}
Efficient Graph processing is challenging because of the irregularity
of graph algorithms. Using GPUs to accelerate irregular graph algorithms 
is even more difficult to be efficient, since GPU's highly structured 
SIMT architecture is not a natural fit for irregular applications. 
With lots of previous efforts spent on subtly mapping graph algorithms
onto the GPU, the performance of graph processing on GPUs is still
highly memory-latency bound, leading to low utilization of compute
resources. Random memory accesses generated by the sparse 
graph data structure are the major causes of this significant 
memory access latency. Simply applying the conventional cache blocking 
technique proposed for matrix computation have limited 
benefit due to the significant overhead on the GPU.

We propose GraphCage, a cache centric optimization framework for 
highly efficient graph processing on GPUs. We first present a 
throughput-oriented cache blocking scheme (TOCAB) in both push 
and pull directions. Comparing with conventional cache blocking
which suffers repeated accesses when processing large graphs on 
GPUs, TOCAB is specifically optimized for the GPU architecture 
to reduce this overhead and improve memory access efficiency. To 
integrate our scheme into state-of-the-art implementations without 
significant overhead, we coordinate TOCAB with load balancing 
strategies by considering the sparsity of subgraphs. To enable cache 
blocking for traversal-based algorithms, we consider the benefit and 
overhead in different iterations with different working set sizes, 
and apply TOCAB for topology-driven kernels in pull direction. 
Evaluation shows that GraphCage can improve 
performance by 2 $\sim$ 4$\times$ compared to hand optimized 
implementations and state-of-the-art frameworks (e.g. CuSha 
and Gunrock), with less memory consumption than CuSha.
\end{abstract}

\maketitle
\section{Introduction}
Analytics and knowledge extraction on graph data structures have 
become areas of great interest in today's large-scale datacenters, 
as social network analysis and machine learning applications 
have received considerable attention. Many software systems 
~\cite{Pregel,Graphlab,DisGraphlab,Gemini} have been developed for 
this application domain. Graph algorithms are usually embarrassingly 
parallel, which makes massively parallel accelerators (e.g. GPUs) 
possible candidates to speedup them. However, efficient graph processing
on GPUs is challenging because 1) graph algorithms are irregular and 2) GPUs 
have highly structured architecture. It is difficult to avoid inefficiency when 
mapping irregular algorithms to a structured architecture. Specifically there 
are four major problems for GPU graph processing: (1) low work efficiency, (2) 
high synchronization overhead, (3) load imbalance and (4) poor data locality. 

Lots of previous work have contributed to alleviate the effects 
of the first three problems, but the poor locality problem has 
got little attention on GPUs. This is because 1) with thousands 
of threads running simultaneously, GPUs have relatively much 
smaller cache capacity per thread compared to CPUs, which makes 
cache oriented optimization techniques less effective on GPUs,
2) GPU memory hierarchy is quite different from that of CPU,
which makes CPU cache optimization techniques not directly
applicable to the GPU, 3) GPU L2 cache is usually used as a
global synchronization point instead of holding the working set, 
and GPU programmers tend to use scratchpad for locality issues,
and 4) it is hard to capture locality from random accesses in graph 
algorithms. Fortunately, recent GPUs have shown a trend of including 
larger caches on chip, making cache oriented optimization a more 
important role in achieving highly efficient graph processing on GPUs.

In this work, we focus on improving cache performance for graph
processing on GPUs. Graph algorithms have poor cache utilization
on GPUs due to the random memory accesses to the vertex values. 
When the graph is large enough, random accesses cause frequent 
cache misses and long memory access latency. Cache blocking is an 
optimization technique to improve locality by partitioning a data
structure into small blocks such that each block can fit in cache.
CuSha~\cite{CuSha} is a GPU graph processing framework that uses
the idea of cache blocking. It partitions input graphs into 
\textit{Shards} to fit into the GPU shared memory. This approach 
leverages fast shared memory in GPUs, but limited by shared memory 
size it produces too many subgraphs which leads to non-trivial overhead. 
Meanwhile, CuSha uses a COO-like graph representation that requires 
roughly 2.5 $\times$ global memory space than the typical CSR format.
To overcome the limitations, we propose to leverage the last level 
cache (LLC) in the GPU to improve locality. Meanwhile, the graph 
representation is based on CSR to save global memory space.

However, naively applying cache blocking on LLC can bring a large volume 
of inefficient memory accesses on the GPU. In this paper, we design 
a static cache blocking scheme for both pull and push directions.
Our scheme replaces sparse accesses to the global sums with dense 
accesses to partial sums, which improves memory access efficiency. 
We also coordinate our optimization with state-of-the-art load balancing. 
For traversal-based graph algorithms, we consider the iterations in 
different phases according to their dynamic change of the working set size, 
and apply our cache blocking scheme for topology-driven kernels in pull direction.
We implement a graph processing framework GraphCage that integrates our
proposed optimizations. We evaluate GraphCage on a recent NVIDIA GPU with 
diverse real-world graphs. Experimental results show that GraphCage 
achieves speedup of 2$\times$ over hand optimized implementations,
and 4$\times$ over the implementations in state-of-the-art GPU graph
processing frameworks Gunrock. Compared to a previous GPU graph 
partitioning framework, CuSha, GraphCage achieves better performance
with less memory consumption.

This paper makes the following contributions:

\begin{itemize}
\item We investigate the cache performance of state-of-the-art graph
algorithms on GPUs and observe poor cache utilization caused by 
random memory accesses.

\item We propose GraphCage, a cache aware optimization framework
for efficient GPU graph processing. The core is
a static cache blocking scheme specialized for the GPU.

\item We apply our cache optimization with load balancing, and enable 
it for traversal-based algorithms, to achieve high performance for 
different applications.

\item We measure GraphCage on the NVIDIA GPU and show
superior performance over state-of-the-art implementations.
\end{itemize}

The rest of the paper is organized as follows: Section~\ref{sect:back} 
introduces the background and explains our motivation.
The GraphCage design is described in Section~\ref{sect:design}.
We present the evaluation in Section~\ref{sect:eval}.
Section~\ref{sect:relate} summarizes related works and
Section~\ref{sect:concl} concludes.

\section{Background and Motivation}\label{sect:back}
Recently people have intensively investigated many graph 
algorithms on various platforms, 
and proposed plenty of parallel strategies and optimization 
techniques to make them as fast as possible. Meanwhile, many 
graph processing frameworks and libraries have been developed 
to improve the performance and programmability of graph applications 
targeting CPUs~\cite{Ligra,Galois,GraphMat,GraphReduce,GraphPad} 
and GPUs~\cite{Gunrock,MapGraph,Medusa}, by generalizing 
these proposed optimizations. In this section, we first 
summarize these strategies and optimizations. We then point 
out that there is still room for performance improvement 
even if these sophisticated optimizations have been applied. 
Finally we briefly introduce cache blocking, and show 
its limitations when applied to GPU graph processing.

\subsection{Graph Processing on GPUs}
\begin{figure}[t]
\begin{center}
	\includegraphics[width=0.5\textwidth]{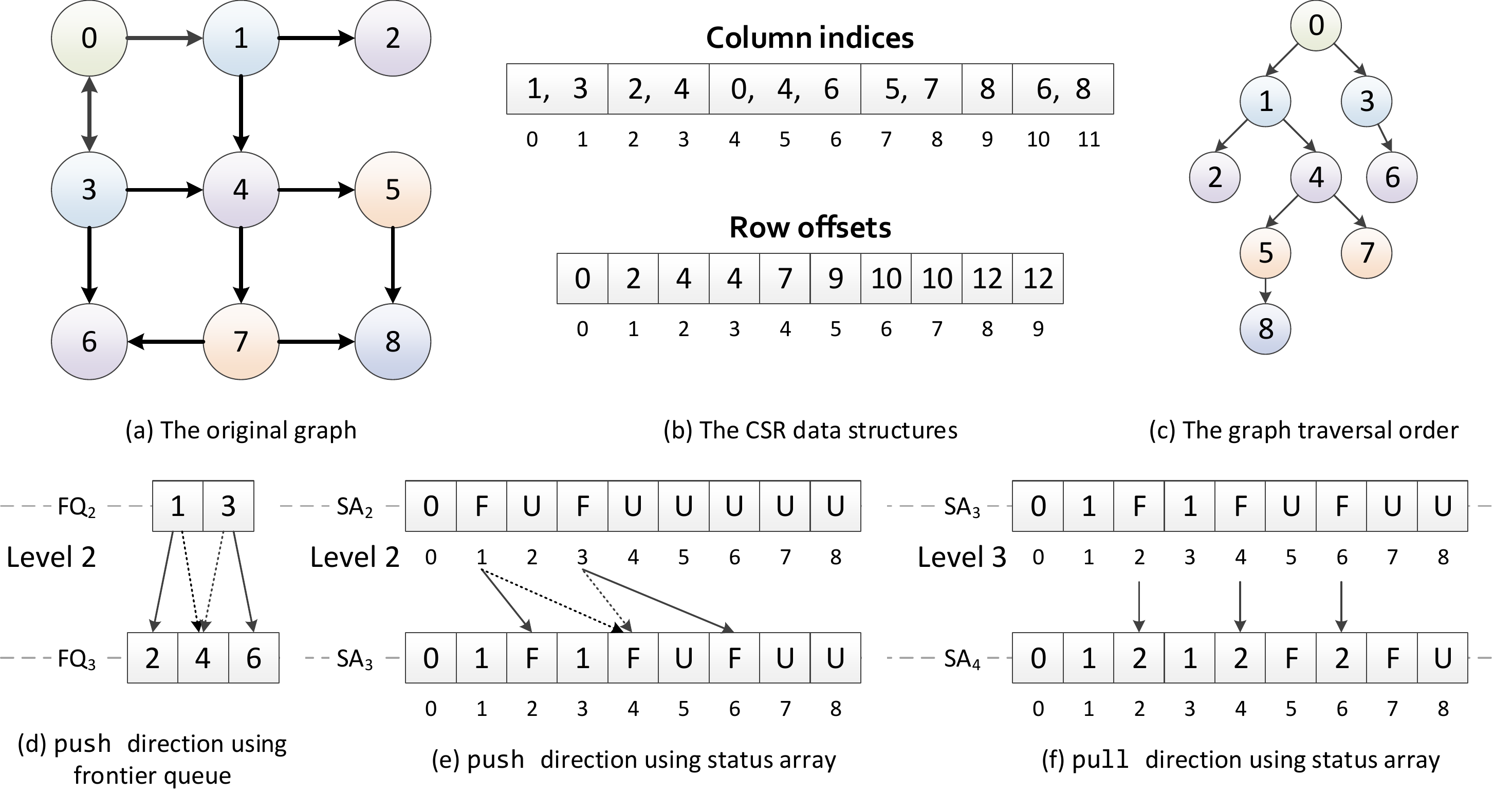}
	\vspace{-0.6cm}
	\caption{An example of the CSR data structure, frontier queue and status array.  
		The number in the status array represents the depth of each vertex.
		The labels of F and U represent frontier and unvisited vertex respectively.}
	\vspace{-0.6cm}
	\label{fig:example}
\end{center}
\end{figure}

Efficient graph processing on GPUs is notoriously difficult. 
There has been considerable recent effort in improving the 
performance of various graphs algorithms on both CPUs and GPUs. 
Generally, there are four key issues that affect the performance most:
1) work efficiency~\cite{Direction,DVT,Luo,Hong,Grossman,CC}, 
2) load balancing~\cite{Merrill,SSSP,GraphGrind,Enterprise}, 
3) synchronization overhead~\cite{Merrill,PushPull}, and 
4) data locality~\cite{Beamer,Zhang,PropBlocking,SpMV}. 
Both algorithm-level and architecture-oriented optimization techniques 
have been proposed targeting one or several of these issues. In this section, 
we briefly summarize these methods and explain how they are applied on GPUs. 
We use ~\cref{fig:example} as an example to illustrate the basic ideas. 
\cref{fig:example} (a) is a directed graph whose CSR format is shown in 
\cref{fig:example} (b).

We use vertex-centric model~\cite{Tao} in this work. 
There are two kinds of graph algorithms: 
1) All-active: all the vertices are active in every iteration, i.e. PageRank. 
2) Partial-active: only some of the vertices are active in each iteration and 
the \textit{active vertex set} is changing during the execution, i.e. BFS. The 
algorithm converges when the active vertex set is empty. Many graph traversal 
based algorithms are partial-active, and the frontier queue in the graph 
traversal is the active vertex set. Optimizations used for all-active and 
partial-active algorithms may be different because the active vertex set could 
substantially change the program characteristics.

\begin{figure}[t]
	\begin{center}
		\includegraphics[width=0.36\textwidth]{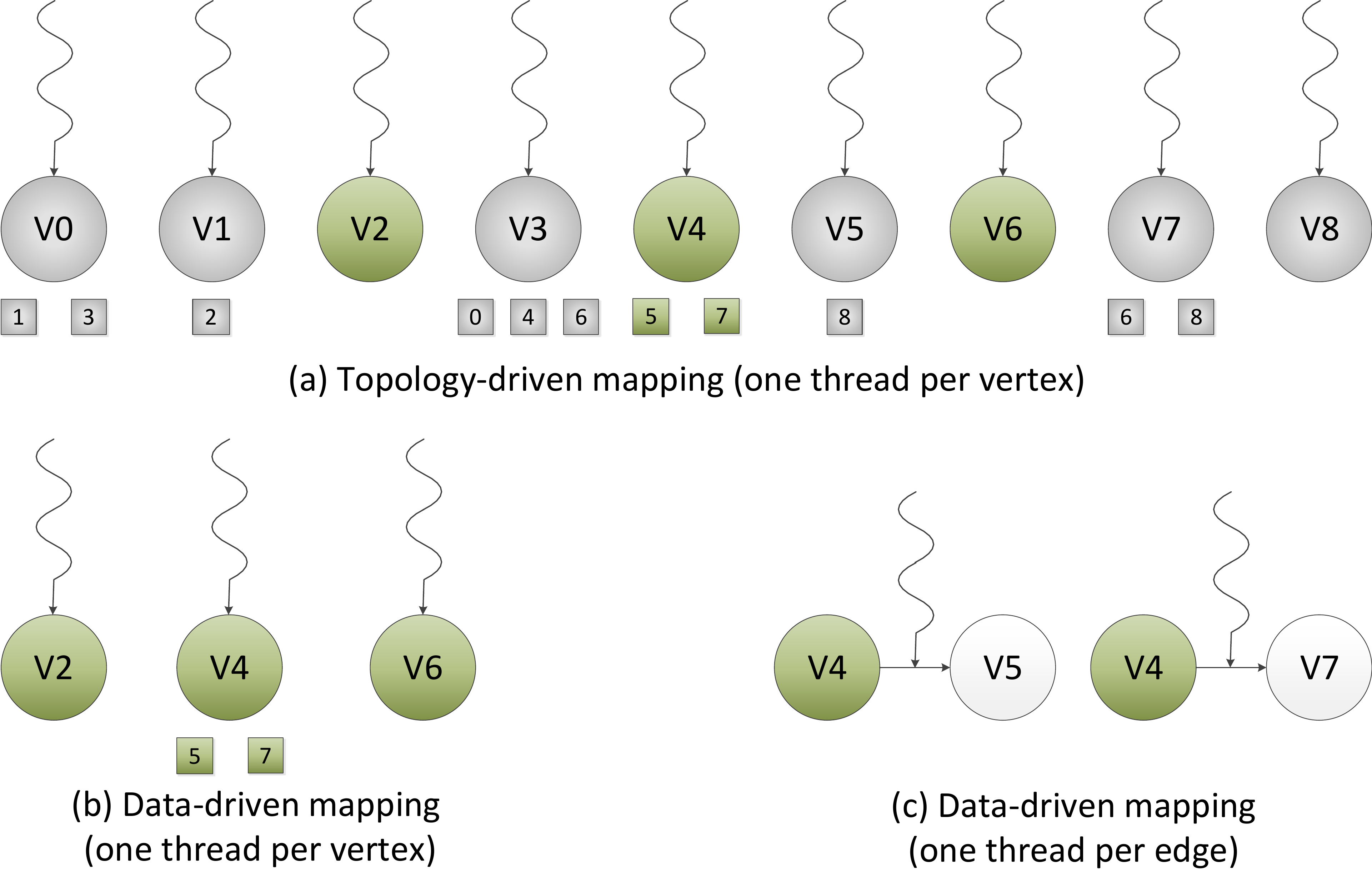}
		\caption{Distribution of threads to units of work.}
		\vspace{-0.5cm}
		\label{fig:we}
	\end{center}
\end{figure}

\textbf{Work Efficiency.}
It is important to improve work efficiency on a parallel machine 
(especially on a massively parallel processor) to take advantage of its 
compute capability. Work efficiency highly depends on how the threads 
are mapped to the tasks. On GPUs, usually two mapping strategies are 
utilized: \textit{topology-driven} or \textit{data-driven}, to map the 
work to the threads~\cite{DVT}. As shown in~\cref{fig:we} (a), the
topology-driven scheme assigns a thread to each vertex in the graph, 
and in each iteration, the thread stays idle or processes a vertex 
depending on whether its corresponding vertex is active or not. 
It is straightforward to implement the topology-driven scheme on GPU  
with no extra data structure. By contrast, the data-driven scheme 
maintains an active vertex queue which holds the vertices to be 
processed during a certain iteration. In this iteration, threads 
are created in proportion to the size of the queue. Each thread 
is responsible for processing a certain amount of vertices in 
the queue and no thread is idle, as shown in \cref{fig:we} (b). 
Therefore, the data-driven scheme is generally more 
work-efficient than the topology-driven one when a small portion 
of the vertices are active, but it needs extra overhead to maintain the 
queue. Mapping strategies usually work with the \textit{direction 
optimization}~\cite{Direction} technique to improve work efficiency. 
Graph algorithms can be implemented in either pull or push direction. 
\cref{fig:example} (c) illustrates the traversal order when performing 
breadth first search (BFS) on the graph in \cref{fig:example} (a). 
In level 2 we apply push direction with either a frontier queue 
(\cref{fig:example} (d)) or a status array (\cref{fig:example} (e)). 
In level 3 pull direction is used (\cref{fig:example} (f)). 
This hybrid method dynamically changes the direction to maximize work 
efficiency for each iteration (level).

\textbf{Load Balancing.}
Since vertices have different amount of edges to be processed by the 
corresponding threads, load imbalance becomes an important issue for 
vertex-centric graph processing, particularly when the input is a scale-free 
graph in which neighbor sizes of vertices can differ by orders of magnitude.  
Hong~\emph{et~al.}~\cite{Hong} proposed a \textit{Virtual Warp Centric} (VWC) 
method for BFS to map warps rather than threads to vertices. Since the edges of 
a vertex are stored continuously, this approach can not only balance the work 
inside the warp, but also lead to coalesced memory accesses, which improves 
bandwidth efficiency. As an extension, Merrill ~\emph{et~al.}~\cite{Merrill} 
proposed a hierarchical load balancing strategy \textit{Thread+warp+CTA} (TWC) 
which maps the workload of a single vertex to a thread, a warp, or a thread 
block, according to the size of its neighbor list. At the fine-grained level, 
all the neighbor list offsets in the same thread block are loaded on chip, 
then the threads in the block cooperatively process per-edge operations. 
At the coarse-grained level, per-block and per-warp schemes are utilized to 
handle the extreme cases: (1) neighbor lists larger than a thread block; (2) 
neighbor lists larger than a warp but smaller than a thread block respectively. 
\cref{fig:we} (c) shows a load balanced data-driven mapping, where each thread 
is assigned to process an edge. In this way, the TWC approach can handle 
vertices with different sizes of neighbor lists. Enterprise ~\cite{Enterprise} 
further improves TWC with inter-block load balancing for handling extremely 
high-degree vertices (hub vertices) using the entire grid. 

\textbf{Synchronization overhead.}
Updating vertex values in parallel may cause conflicts, which requires 
atomic operations to guarantee correctness. Usually pull-style 
operations do not need atomics since vertex values are accumulated by 
individual threads separately. However, conflicts happen for push-style 
operations as different threads could write to the same vertex value 
concurrently. Therefore, for those all-active algorithms, such as PR, 
pull is usually faster than push as it requires no atomic operation.  
But when we apply fine-grained load balancing scheme, such as TWC, 
to pull-style algorithms, conflicts occur among cooperative threads 
in the thread block, and thus it brings in atomic operations. 
This leads to significant synchronization overhead which is not 
acceptable. So usually we only apply coarse-grained load balancing 
in pull direction. On the contrary, TWC works well for push-style 
operations as synchronization can not be avoided no matter what 
strategy is used. Besides, maintaining the frontier queue in partial-active algorithms also requires synchronization~\cite{Merrill}. 

\textbf{Data locality.}
Most of the above mentioned optimizations have been incorporated 
in state-of-the-art graph processing frameworks such as Gunrock 
~\cite{Gunrock}. Therefore, graph applications written within the 
Gunrock framework can achieve excellent performance close to that 
of hand-optimized implementations. By contrast, CuSha~\cite{CuSha} 
is a framework mainly focusing on partitioning the input graph into 
\textit{shards} and therefore can coalesce memory accesses and 
make good use of the scratchpad. This method is very similar to the 
\textit{propagation blocking} technique used in CPU graph algorithms 
~\cite{PropBlocking,SpMV}. CuSha can also improve performance 
compared to non-partitioned implementations. However, since the subgraph 
size is confined by the size of scratchpad in each SM, each subgraph is 
small and there are too many blocks generated. This leads to non-trivial 
overhead on merging partial results of subgraphs. 

\subsection{Performance Bottleneck on GPUs}
The parallel strategies and optimization techniques introduced 
above have substantially improved the performance of graph
processing on GPUs~\cite{Gunrock}. However, the GPU hardware
is still poorly utilized. We conduct a characterization on the 
GPU using hand optimized PageRank. We observe low hardware 
utilization, including both the compute units and memory bandwidth.
This result indicates that the performance is still limited by memory 
access latency, even if state-of-the-art optimizations are applied. 

\begin{algorithm}[t]
\caption{PageRank in pull direction}
\label{alg:pull}
\begin{algorithmic}[1]
	\Procedure{Pull}{$G^T$($V$,$E$)}
		\For{each vertex $u \in V$ \textbf{in parallel}}
			\State contributions[$u$] $\leftarrow$ rank[$u$]/out\_degree[$u$]
		\EndFor
		\For{each vertex $dst \in V$ \textbf{in parallel}}
			\State sum $\leftarrow$ 0
			\For{each incoming neighbor $src$ of vertex $dst$}
				\State sum $\leftarrow$ sum + contributions[$src$]
			\EndFor
			\State sums[$dst$] $\leftarrow$ sum
		\EndFor
		\For{each vertex $u \in V$ \textbf{in parallel}}
			\State rank[$u$] $\leftarrow$ (1-d)/$|V|$ + d $\times$ sums[$u$]
		\EndFor
	\EndProcedure
\end{algorithmic}
\end{algorithm} 

\begin{algorithm}[t]
	\caption{PageRank in push direction}
	\label{alg:push}
	\begin{algorithmic}[1]
		\Procedure{Push}{$G$($V$,$E$)}
		\For{each vertex $u \in V$ \textbf{in parallel}}
			\State contributions[$u$] $\leftarrow$ rank[$u$] / out\_degree[$u$]
		\EndFor
		\For{each vertex $src \in V$ \textbf{in parallel}}
			\State contribution $\leftarrow$ contributions[$src$]
			\For{each outgoing neighbor $dst$ of vertex $src$}
				\State atomicAdd(sums[$dst$], contribution)
			\EndFor
		\EndFor
		\For{each vertex $u \in V$ \textbf{in parallel}}
			\State rank[$u$] $\leftarrow$ (1-d)/$|V|$ + d $\times$ sums[$u$]
		\EndFor
		\EndProcedure
	\end{algorithmic}
\end{algorithm}

The inefficiency of the memory hierarchy is due to the large volume of 
the input data and the irregular memory access pattern of the data. 
When executed on a GPU, this large volume of irregular memory 
accesses can not be coalesced. They are translated into a huge 
amount of load and store transactions in the memory hierarchy.
Demand-fetched caches in GPUs have very limited capacity and
are impossible to hold this huge amount of data on chip~\cite{Jia}.
This causes the cache thrashing problem~\cite{Chen}, leading to 
a large number of DRAM accesses. Therefore graph algorithms 
exhibit very poor temporal and spatial locality, although there is 
plenty of potential data reuses exist in these algorithms~\cite{Beamer}. 
Software managed local memory (or scratchpad) is also not 
capable to capture the data reuses for these random accesses.
On the other hand, since most of the memory requests are sent 
to DRAM and cause significant memory access latency, the SM 
pipelines are often stalled waiting for the return of memory requests, 
which leads to inefficiency of the compute units as well.

To understand the performance bottleneck more specifically,
we use PageRank (PR) as an example. PR~\cite{PageRank} is a
graph algorithm for ranking a website based on the score (rank) 
of the sites that link to it. PR is iterative; at each iteration, 
it updates the score of each vertex using the weighted sum 
of its neighbors' scores and degrees. PR can be implemented 
in either pull or push direction. Algorithm~\ref{alg:pull} shows 
one iteration of pull-based PR (PR-pull), while Algorithm
~\ref{alg:push} is in push style (PR-push). For each destination vertex, 
PR-pull gathers contributions from all the incoming neighbors 
(line 6 \& 7) and accumulate them to \texttt{sums} (line 8).  
On the contrary, PR-push scatters the contribution of each source 
vertex to all its outgoing neighbors (line 6). Therefore, when 
implemented in parallel, atomic operations are required (line 7).

For both algorithms, lines from 4 to 7 are the most important, which 
consume most of the execution time. PR-pull avoids atomic operations, 
and the sum of incoming contributions will have high locality. But the 
random memory accesses to the \texttt{contributions} of source vertices 
(line 7 in Algorithm~\ref{alg:pull}) could have low locality and become 
the major performance limiter. PR-push requires atomic operations, and 
therefore is usually slower than PR-pull. In push style, the outgoing 
contribution will have high locality, but reading (and writing) its 
neighbors' \texttt{sums} (destination vertices) could have low locality 
(line 7 in Algorithm~\ref{alg:push}), which also causes significant 
memory latency and limits overall performance.

\begin{algorithm}[t]
	\caption{Shortest Path Calculation in Betweenness Centrality}
	\label{alg:bc_forward}
	\begin{algorithmic}[1]
		\Procedure{BC\_Forward}{$G$($V$,$E$)}
		\For{each vertex $src \in Q_{curr}$ \textbf{in parallel}}
			\For{each outgoing neighbor $dst$ of vertex $src$}
				\If{atomicCAS($\delta$[$dst$], $\infty$, $\delta$[$src$]+1) = $\infty$}
					\State $Q_{next}$.push($dst$)
				\EndIf
				\If{$\delta$[$dst$] = $\delta$[$src$] + 1}
					\State atomicAdd($\sigma$[$dst$],$\sigma$[$src$])
				\EndIf
			\EndFor
		\EndFor
		\EndProcedure
	\end{algorithmic}
\end{algorithm}

We use another example, Betweenness Centrality (BC)~\cite{BC}, to show 
the characteristics of traversal based algorithms such as BFS and SSSP. 
BC is commonly used in social network analysis to measure the influence 
a vertex has on a graph. It calculates a betweenness centrality 
\textit{score} for each vertex in a graph based on shortest paths. For 
every pair of vertices in a connected graph there exists at least one 
shortest path between them.  A vertex's betweenness centrality score is 
related to the fraction of shortest paths between all vertices that pass 
through the vertex. A push style data-driven shortest path calculation 
kernel in BC is shown in Algorithm~\ref{alg:bc_forward}, where $\delta$ is 
the depth and $\sigma$ is the number of shortest paths for 
each vertex. The parallel for loop in Line 2 assigns one thread to each 
element in the frontier queue such that only active vertices in each 
iteration are traversed. The atomic Compare and Swap (CAS) operation 
on Line 4 is used to prevent multiple insertions of the same vertex into 
$Q_{next}$. Since this kernel is in push style, it requires atomics to update
$\sigma$ (line 7). In this kernel, random accesses to $\delta$ and 
$\sigma$ of the destination vertices cause cache thrashing.

\subsection{Cache Blocking}
Although random accesses can not be avoided anyway, there is still plenty 
of room for performance improvement if we can make good use of caches by 
reordering the memory accesses. Cache blocking~\cite{blocking,CSB} is 
such a reordering technique proposed to improve cache performance for 
dense linear algebra and other applications. By partitioning the graph 
into small blocks, the range of vertex values in each subgraph is 
reduced significantly such that the corresponding the arrays data 
for vertex values are small enough to reside in cache, therefore 
improving the locality~\cite{PropBlocking}. Cache blocking can be done 
in 1D or 2D, and can be applied to either push or pull direction. 
Although cache blocking can significantly improve performance for dense 
matrix computation, naively applying cache blocking to graph processing 
on GPUs is much less effective. Take PR-pull as an example. Cache 
blocking for PR-pull divides the adjacency matrix into column blocks.
This will improve the locality of reading the neighbors' contributions 
(line 7 in Algorithm~\ref{alg:pull}), as the source vertex id (the index 
of the \texttt{contributions} array) is restricted in the partition, 
and therefore the corresponding vertex values can fit in cache. But column 
blocking results in repeated accesses to \texttt{sums} (line 8 in Algorithm 
~\ref{alg:pull}) for each block. In another word, cache blocking reduces 
accesses to one data structure at the cost of increasing accesses to another 
data structure. 

Actually, in each subgraph, the number of edges is much fewer than that in 
the original graph. Many destination vertices may not have edge in a given 
subgraph since source vertices of this subgraph are restricted in a small range. 
For these vertices, accessing \texttt{sums} is useless. However, for GPU
implementations, even if there is only one thread in a warp has updated 
\texttt{sums}, there will be a memory write request generated for this update. 
This memory access pattern is inefficient spatially. When the graph is not 
very large and it is partitioned into only several subgraphs, these inefficient 
memory accesses may not be a problem. But when a graph is large enough to have 
tens of subgraphs or more, this overhead becomes non-trivial and cache blocking 
achieves only marginal performance improvement or could even lead to slowdowns.

\section{GraphCage Design}\label{sect:design}
In this section, we describe our GraphCage design in a progressive way. 
We first propose a throughput-oriented cache blocking (TOCAB) 
scheme which is orchestrated specifically for the GPU architecture. 
TOCAB is applicable for both pull and push style algorithms.
We then integrate TOCAB with load balancing to achieve high performance.
Finally, we enable cache blocking for traversal based algorithms 
by considering the change of working set size.

\subsection{Throughput Oriented Cache Blocking}
As aforementioned, conventional cache blocking for CPUs is much less
effective for GPUs. To enable efficient cache blocking on GPUs, we 
propose Throughput Oriented Cache Blocking (TOCAB) that is designed 
for the GPU architecture. At the beginning, we have several design 
choices to make. First, although modern GPUs have multiple levels of 
caches with different types (e.g. read-only cache and software managed 
scratchpad memory), in this work our cache blocking scheme mainly 
focuses on the last-level cache (LLC), i.e. the L2 cache in most of 
the modern GPU architectures. This is reasonable because 1) scratchpads 
or read-only caches are level-one caches with relatively small capacities, 
which severely limits the block size of cache blocking; 
2) the latency and bandwidth between the LLC and DRAM is typically the 
performance bottleneck and we are committed to reduce the overall 
number of DRAM accesses. Note that scratchpad and read-only caches are 
also useful and we will discuss how to leverage them for locality later.

Second, we apply one dimensional (1D) instead of two dimensional (2D) 
blocking. This is because 1) transforming the CSR data structure is  
easier for 1D blocking, which means less preprocessing overhead; 2) 2D 
blocking partitions the graph on both source and destination vertices, 
which may generate too many small blocks in total. Smaller blocks result 
in fewer data reuses to capture and less parallelism for the GPU to 
exploit, while more blocks lead to more overhead for merging the partial 
results of each block. When there are too many blocks, the overhead may 
outweigh the benefit of cache blocking and therefore leads to limited 
performance improvement or even slowdown. In this case, the block size 
is actually a tradeoff between locality benefit and blocking overhead. 
To obtain a scalable block size, we decide to use 1D blocking.

\begin{figure}[t]
	\begin{center}
		\includegraphics[width=0.4\textwidth]{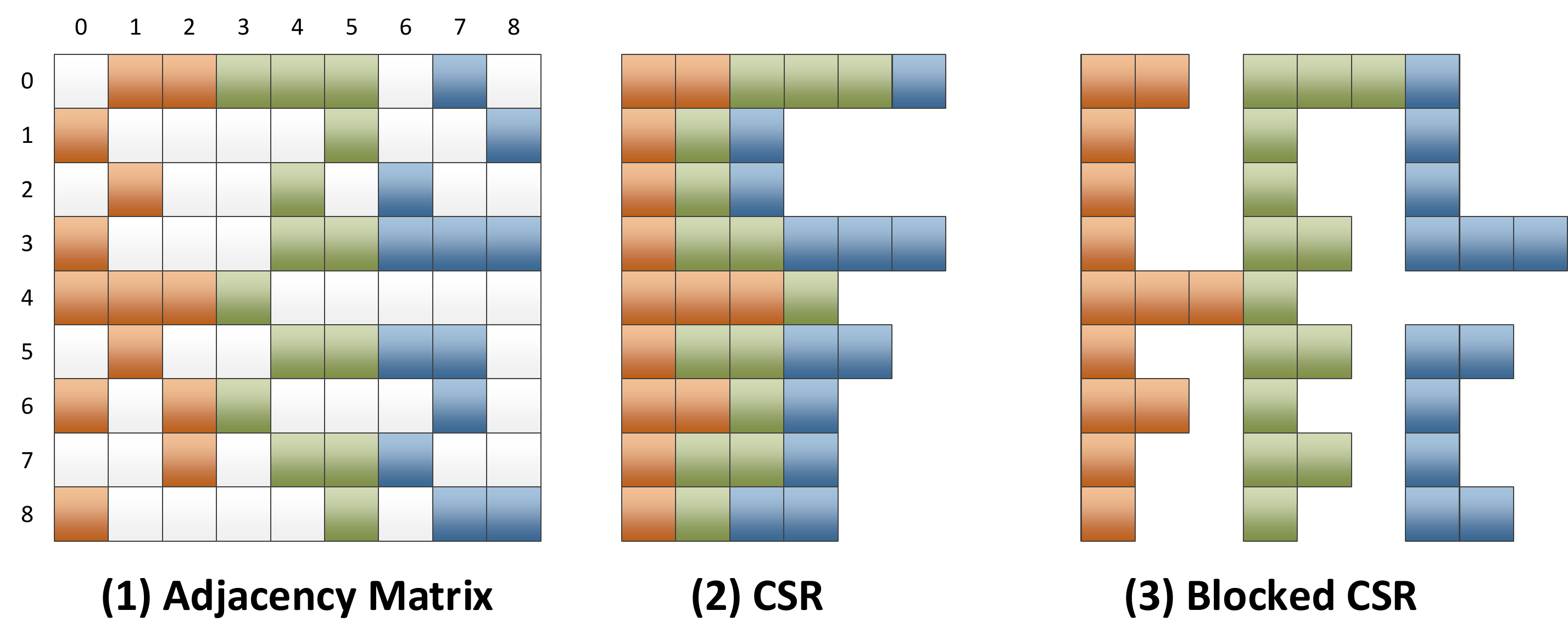}
		\caption{The CSR representation before and after cache blocking (pull-style).}
		\vspace{-0.5cm}
		\label{fig:blockedCSR}
	\end{center}
\end{figure}

Third, we choose to use static blocking instead of dynamic blocking. 
Static blocking divides the graph into subgraphs before the execution of 
the algorithm, and therefore incurs less runtime overhead. By contrary, 
dynamic blocking~\cite{PropBlocking,milk} uses intermediate buffers to 
hold the data at runtime, and the data is dynamically distributed 
into different buffers so that each buffer can fit in cache. The 
intermediate data is processed separately and then the partial results 
are accumulated. Dynamic blocking requires no or little modification on 
the original CSR data structure, but needs a huge memory space for the 
intermediate buffers. In addition to consuming more memory, 
dynamic blocking performs additional load and store instructions in 
order to insert data into the buffers and read data back from the 
buffers. Note that preprocessing can be applied to reduce this runtime 
overhead of dynamic blocking, which is a tradeoff between static and 
dynamic overhead. Anyway, preprocessing (static overhead) is necessary 
to ensure high performance for both static and dynamic approaches. We 
choose static blocking to reduce as much dynamic overhead as possible, 
because 1) most of graph algorithms are iterative and take a large 
number of iterations, making a case for additional preprocessing time; 
2) preprocessing cost is small compared to the significant performance 
gains; 3) the partitioned graphs can also be reused across multiple 
graph applications, further amortizing the preprocessing cost; 4) the 
dynamic operations are simpler after the graph is statically blocked, 
making it easier to implement on GPUs for different graph applications.

Based on these design choices, TOCAB can be divided into three phases. 
First, the original CSR data structure is modified to store the 
subgraphs, i.e. the preprocessing phase.
Second, each subgraph is processed sequentially, i.e. the 
subgraph processing phase. Third, the partial results of all subgraphs 
are reduced to the final results, i.e. the post-processing (reduction) 
phase. We describe each phase in details as follows.

\begin{algorithm}[t]
	\caption{Subgraph Processing in pull-style PageRank}
	\label{alg:pullb}
	\begin{algorithmic}[1]
		\Procedure{Pull}{$G_{sub}^T$($V_{sub}$,$E_{sub}$)}
		\For{each vertex $dst_{local} \in V_{sub}$ \textbf{in parallel}}
			\State sum $\leftarrow$ 0
			\For{each incoming neighbor $src$ of $dst_{local}$}
				\State sum $\leftarrow$ sum + contributions[$src$]
			\EndFor
			\State partial\_sums[$dst_{local}$] $\leftarrow$ sum
		\EndFor
		\EndProcedure
	\end{algorithmic}
\end{algorithm}

In the preprocessing phase, we partition the graph into subgraphs so that 
the vertex values in each subgraph can fit in cache. One technique is to 
store the subgraphs each as their own graph, which is known as blocked CSR (as shown in \cref{fig:blockedCSR}). For each subgraph, it has its own CSR data structures (e.g. $rowptr$ and $colidx$). Computationally, it is easy to generate blocked CSR for each subgraph. For pull model, we do column blocking which takes each edge in the graph and classifies it to the subgraph which its source vertex belongs to. For push model, destination vertex is used instead of source vertex, and therefore we apply row blocking. Note that the code for column blocking (pull) can be reused for row blocking (push), since the input graph of the push model is just the transpose graph of that used in the pull model. This means the same preprocessing code works for both push and pull models with no modification.

\begin{algorithm}[t]
	\caption{Subgraph Processing in push-style PageRank}
	\label{alg:pushb}
	\begin{algorithmic}[1]
		\Procedure{Push}{$G_{sub}$($V_{sub}$,$E_{sub}$)}
		\For{each vertex $src_{local} \in V_{sub}$ \textbf{in parallel}}
			\State $src_{global} =$ id\_map[$src_{local}$]
			\State contribution $\leftarrow$ contributions[$src_{global}$]
			\For{each outgoing neighbor $dst$ of $src_{local}$}
				\State atomicAdd(sums[$dst$], contribution)
			\EndFor
		\EndFor
		\EndProcedure
	\end{algorithmic}
\end{algorithm}

\begin{figure}[t]
	\begin{center}
		\includegraphics[width=0.35\textwidth]{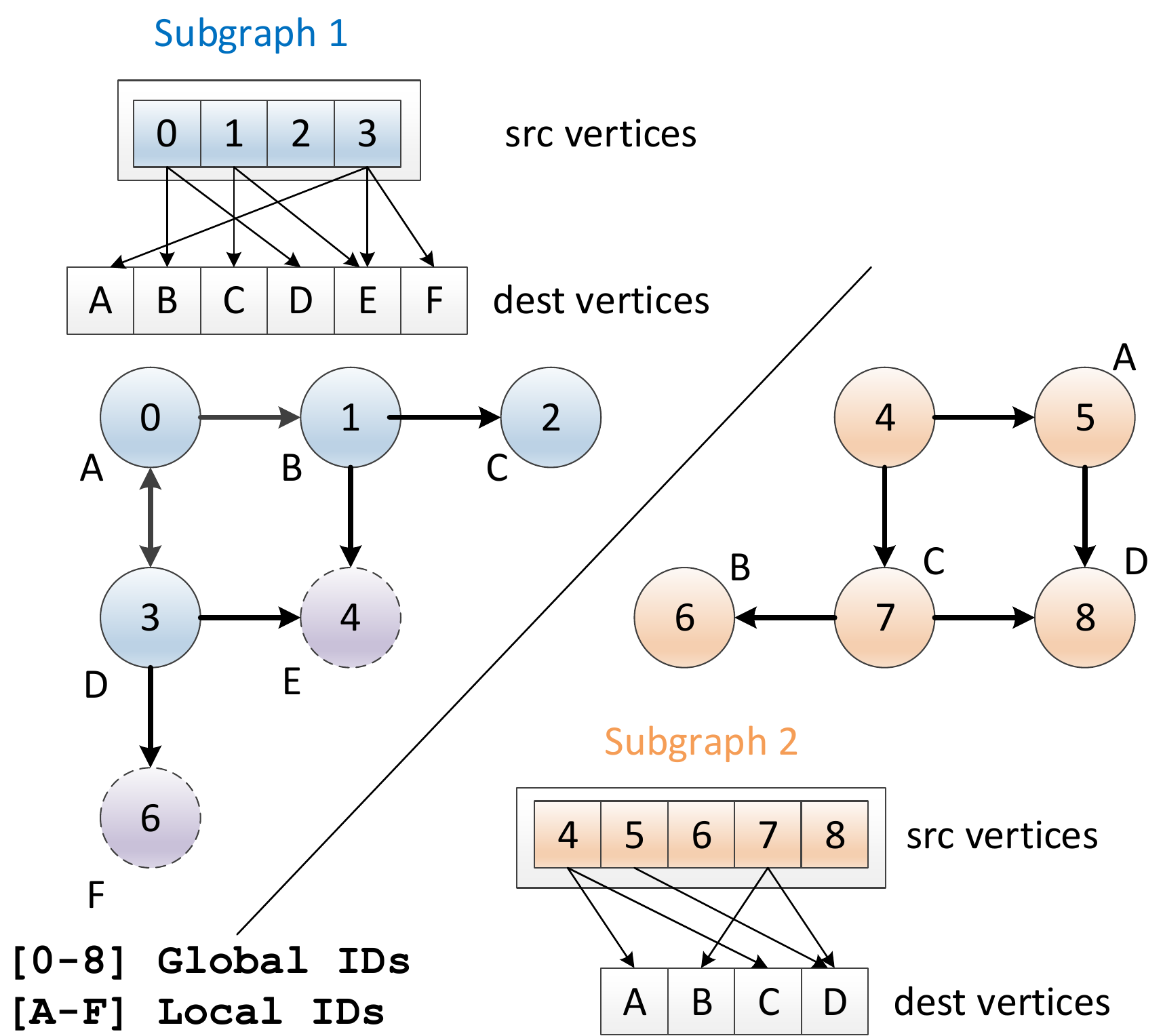}
		\caption{Example of how the graph is partitioned into two 
		subgraphs using source vertices in pull direction. Only 
		vertices that have incoming edges are assigned local-IDs.}
		\vspace{-0.5cm}
		\label{fig:partition}
	\end{center}
\end{figure}

However, blocked CSR is not enough to achieve high performance. As 
mentioned, for pull model, the gain of cache blocking for the low-locality 
access stream (\texttt{contributions}) comes at the expense of the 
high-locality access stream (\texttt{sums}). It is critical to reduce the 
overhead of repeatedly accessing \texttt{sums}. Since only a small portion 
of destination vertices have incoming edges in each subgraph, to improve 
efficiency, we ignore the destination vertices with no edges and assign 
each destination vertices with at least one edge a local-ID. We keep a 
mapping of this local-IDs in the subgraph to the global-IDs used in the 
original graph, so that we can always find the global-ID of a vertex when 
we need to accumulate its partial results. We keep a \texttt{partial\_sums} 
array for each subgraph. The partial results are not directly written into 
\texttt{sums}, but stored in \texttt{partial\_sums}. After all the 
subgraphs are processed, all the \texttt{partial\_sums} are reduced 
to get the correct \texttt{sums}. Since we use local-IDs instead 
of global-IDs for destination vertices, the \texttt{partial\_sums} 
accessed by destination vertices in a given subgraph are stored 
contiguously in memory regardless of their global-IDs. This is 
particularly important for GPUs as we can have coalesced memory accesses 
for the \texttt{partial\_sums}. \cref{fig:partition} illustrate how the 
graph in \cref{fig:example} is partitioned into two subgraphs and the 
global-IDs are mapped to local-IDs for the destination vertices.

With this global to local ID transformation, the subgraph processing 
phase requires little modification on the original pull or push 
operations. Algorithm~\ref{alg:pullb} and ~\ref{alg:pushb} show the 
operations to process a subgraph in pull and push style respectively. 
In Algorithm~\ref{alg:pullb}, destination vertices use local-ID 
so that accesses to \texttt{partial\_sums} are coalesced (line 6). In 
Algorithm~\ref{alg:pushb}, accesses to \texttt{contributions} use 
global-ID (line 4), and thus we use the \texttt{id\_map} to convert 
local-IDs into global-IDs (line 3). Since we confine the destination 
vertices in a small range, the atomic operations on \texttt{sums} happen 
in the cache (line 6), which substantially speedup the entire phase.

\begin{figure}[t]
	\begin{center}
		\includegraphics[width=0.39\textwidth]{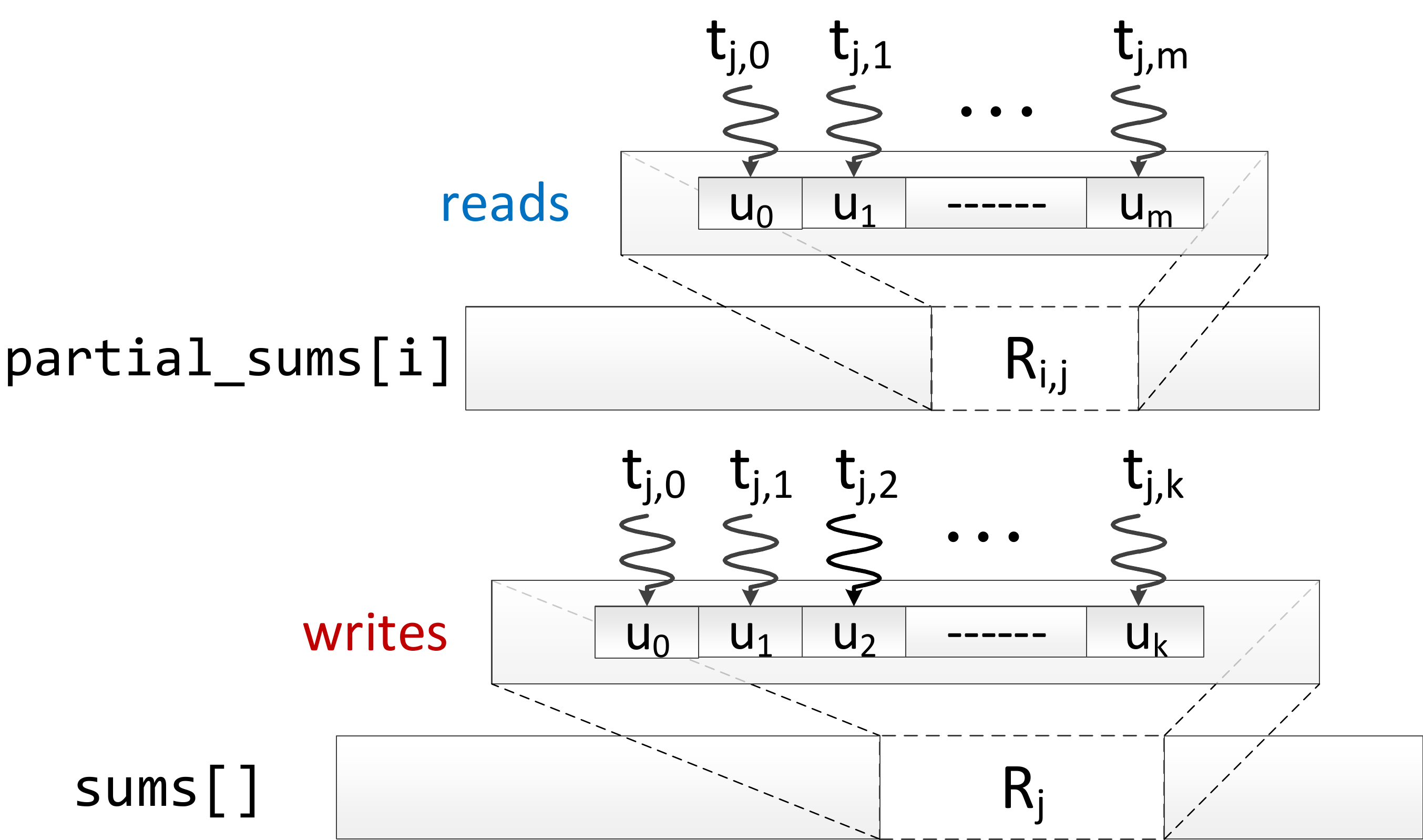}
		\caption{Accumulating \texttt{partial\_sums} into \texttt{sums}.
		$R_{j}$ is the $j$th range. $k$ is the range width.}
		\vspace{-0.3cm}
		\label{fig:merge}
	\end{center}
\end{figure}

In the accumulation phase, only the pull-style implementation needs to reduce 
the partial results, because in push direction the contributions are already 
accumulated into \texttt{sums} as shown in Algorithm~\ref{alg:pullb} line 6. 
We use a GPU friendly approach for accumulation, as shown in \cref{fig:merge}.
Since we use local-IDs in the partial results, we need to find the global
IDs and accumulate the values to the global results (\texttt{sums}).
If different partial results are assigned to different threads, the
write accesses will be inefficient, because the global IDs are not consecutive.
Our strategy is to divide vertices into equal sized ranges (e.g. 1024),
and assign the work of accumulating global results in each range to a thread block.
A thread block is responsible to collect data from the 
specific range of all the subgraphs, and accumulate them in the shared memory. 
When all the partial results are reduced, the final results of this range
are written back to the corresponding position of global memory. In this
case, most of the read and write operations happen in the shared memory, 
while the reads from (\texttt{partial\_sums}) and writes to (\texttt{sums}) 
the global memory are fully coalesced.

\subsection{Coordinated Locality and Load Balancing}
TOCAB improves memory access efficiency so that we have a cache blocking 
scheme that works efficiently on GPUs. To achieve high performance, 
we must integrate TOCAB with other optimization techniques.
When we apply TOCAB with load balancing techniques, we find that cache 
blocking actually affect the load balancing strategy. As aforementioned, 
for PR-push it is easy to apply fine-grained load balancing, since atomic 
operations are not avoidable in push direction. However, we only apply 
coarse-grained load balancing (e.g. VWC) in pull direction to avoid atomics.
Our experiment (not shown in the paper) finds that 
VWC works well for dense-frontier algorithms (e.g PR and SpMV) when the average degree is large (e.g. $\overline{d}>16$), because the memory accesses are consecutive and coalesced, and the SIMD lanes are computationally efficient as most of the threads in a warp are working. 
However, when we partition the graphs into small subgraphs, we find that 
VWC no longer improves performance when applied to each subgraph. In fact, 
the average degree of vertices in the subgraphs are smaller than that in 
the original graph. Table~\ref{table:degree} lists the degree distribution 
of LiveJournal before and after partitioning. In subgraphs more than 90\% 
of the vertices have less than 8 edges and only \%3.5 of the vertices have 
16 or more edges. Since VWC gets each warp to process one vertex and most 
of the SIMD lanes in a warp are idle in this case, making VWC inefficient.

\begin{table}[b]
	\small
	\centering
	\begin{tabular}{c c c}
		\Xhline{2\arrayrulewidth}
		\textbf{degree} & \textbf{Original Graph} & \textbf{Subgraphs}\\
		\hline
		\textbf{0$\sim$7}   & 76.7\% & 90.7\%\\
		\textbf{8$\sim$15}  & 11.7\% & 5.8\% \\
		\textbf{16$\sim$31} & 7.3\%  & 2.3\% \\
		\textbf{32$\sim$}   & 4.3\%  & 1.2\% \\
		\Xhline{2\arrayrulewidth}
	\end{tabular}
	\vspace{0.3cm}
	\caption{Degree distribution of the original graph (LiveJournal) and subgraphs.}
	\label{table:degree}
\end{table}

To coordinate cache blocking with load balancing, we apply different load 
balancing strategies in different directions. For the push-style case, 
whether or not the inner loop is parallelized, atomics are required. 
So we simply apply TWC in push direction. However, for the pull pattern, 
parallelizing the inner loop leads to synchronization overhead~\cite{Grossman}. 
VWC is also not suitable due to its low SIMD efficiency. Therefore, we only 
apply coarse-grained load balancing in pull direction. This is a rational 
choice because hub vertices are still processed by thread blocks or warps, 
while low-degree vertices are processed in serial since in most cases they 
have only a small number of edges and may not be the performance bottleneck.
For some graphs whose performance highly relies on fine-grained load balancing,
using this approach may cause some performance loss. But our evaluation
shows that in general it achieves very good performance for most of the graphs.

\subsection{Work Efficient Cache Optimization}
We have discussed our approach for all-active graph algorithms such as PR. 
However, there are many partial-active graph algorithms, for example, traversal 
based graph algorithms, such as BFS, BC and SSSP. These algorithms have 
different characteristics. In each iteration the operator may update the 
values of only a small subset of vertices. For example, in the first iteration 
of BFS, only the neighbors of the source vertex have their values updated. 
Therefore data-driven mapping with a frontier queue is used to avoid 
unnecessary updates. These differences introduce two extra issues when 
considering cache blocking: 1) the maintenance of the frontier; 
2) the change of working set size.

For a data-driven implementation, when a graph is partitioned into subgraphs, 
extra operations are required to synchronize the frontier queue, which could 
be expensive. Another approach is to use topology-driven mapping with status 
arrays (size of $|V|$) instead of queues, as shown in ~\cref{fig:example} (e). 
Array \texttt{front} is used to show whether the vertices are in the current 
frontier or not. Array \texttt{next} is used to label the vertices in the 
frontier of the next level. When we partition the graph, we assign each 
subgraph a local \texttt{next} array. When processing each subgraph, new 
items are pushed into local status array. After all subgraphs are processed, 
the local status array are reduced into a global status array. This is a 
similar approach used for the \texttt{partial\_sum} in PR. So we can perform 
the reduction of partial results and \texttt{next} in the same kernel.
In this way, the overhead of maintaining activeness of vertices is trivial.

Since the frontier size is changing in different iterations, the size of the 
working set is also changing at runtime. At the beginning, there is only one 
active vertex, the \texttt{source}, in the frontier. And then the queue gets 
larger as more and more vertices are pushed into the frontier. After several 
iterations, the queue starts to shrink until it is empty. The working set 
size is proportional to the total degree of vertices in the frontier. When 
the frontier is small, the working set can possibly fit in cache. It is 
only when the frontier is large enough that cache blocking becomes beneficial.

As direction optimization~\cite{Direction} divides an algorithm into different 
phases, we also use this hybrid method to enable cache blocking for traversal 
based algorithms. Take BC as an example. Algorithm~\ref{alg:bc_forward} is 
split into three phases: push-pull-push. In the two push phases, the working 
set is usually smaller than the cache size, which means cache blocking may not 
be necessary. So we only apply TOCAB in the pull direction. Since the push 
kernel uses data-driven mapping to achieve work efficiency, this choice also 
avoids maintaining the frontier queue locally. For the pull kernel, applying 
TOCAB is the same as PR except updating \texttt{next} array. There are some 
iterations that run in push direction but have a working set larger than the 
cache capacity. Our experiment shows that these large push kernels are rare 
(usually one in each graph). Since the total graph size is confined by the 
memory size, the assumption is reasonable.

Finally we have a full solution for different kinds of graph algorithms.
We integrate the optimizations into a unified framework, GraphCage. Programmers only write basic pull and push kernels
to describe the operators for a specific algorithm, and GraphCage can apply cache optimizations to achieve high performance.
\section{Evaluation}\label{sect:eval}
We evaluate performance of GraphCage on NVIDIA GTX 1080Ti with CUDA 9.1.
The GPU has 3584 cores with boost clock of 1.6GHz.
It has a 2.75 MB L2 cache and 11GB GDDR5X memory with 484 GB/s bandwidth.
We use \texttt{gcc} and \texttt{nvcc} with the \texttt{-O3} 
optimization option for compilation along with \texttt{-arch=sm\_61} 
when compiling for the GPU. We execute all the benchmarks 10 times 
and collect the average execution time to avoid system noise. 
Timing is only performed on the computation part of each program. 

We select our datasets from the UF Sparse Matrix Collection
~\cite{Florida}, the SNAP dataset Collection~\cite{SNAP}, 
and the Koblenz Network Collection~\cite{Konect} which are 
all publicly available. Size, density, topology and 
application domains vary among these graphs. Note that graph 
layout has a tremendous impact on the locality of the vertex 
value accesses~\cite{PropBlocking}. In another word, some 
graphs can benefit from locality optimizations, while some 
others may not, depending on their data layouts. The fact is 
that, for some graphs (e.g. social networks), their topologies 
make it difficult to find a good layout for high locality. 
Meanwhile, for some situations the time spent on transforming 
the graph into a locality-friendly layout is not warranted~\cite{PropBlocking}. 
Therefore, our proposed method is useful in these cases. 
Our experiment mainly shows the performance improvement on 
graphs with poor locality, while we also show the performance 
effect on those graphs with good locality (\texttt{Hollywood}) 
to indicate that GraphCage only causes trivial slowdown.

\begin{table}[t]
	\footnotesize
	\centering
	\resizebox{0.5\textwidth}{!}{
		\begin{tabular}{c c c c c c c}
			\Xhline{2\arrayrulewidth}
			\bf{Graph} & \bf{\# V} & \bf{\# E} & \bf{$\overline{d}$} & \bf{Description}\\
			\hline
			\texttt{LiveJ} & 4.8M & 68M & 14.2 & LiveJournal social network\\
			\texttt{Wiki2007} & 3.6M & 45M & 12.6 & Links in Wikipedia pages\\
			\texttt{Flickr} & 2.3M & 33M & 14.4 & Flickr user connections\\
			\texttt{Wiki-link} & 12M & 378M & 31.1 & Large Wikipedia links\\
			\texttt{Hollywood*} & 1.1M & 112M & 98.9 & Movie actor network\\
			\texttt{Kron21*} & 2M & 182M & 86.8 & Synthetic power-law graph\\
			\texttt{Orkut*} & 3M & 212M & 71.0 & Orkut social network\\
			\texttt{Twitter*} & 21M & 530M & 24.9 & Twitter user connections\\
			\Xhline{2\arrayrulewidth}
		\end{tabular}
	}
	\caption{Suite of benchmark graphs. * indicates that the original graph is undirected. $\overline{d}$ is the average degree.}
	\vspace{-0.5cm}
	\label{table:bench}
\end{table}

In this paper, we use two graph algorithms, i.e. PageRank 
(PR) and Betweenness Centrality (BC) as graph benchmarks.
PR is non-traversal based and all the vertices in every 
iteration are active, while BC is a traversal based algorithm 
whose active vertex set (frontier queue) changes in each 
iteration during the execution. We also include sparse matrix 
vector multiplication (SpMV)~\cite{SpMVGPU} as another 
benchmark since most of graph algorithms can be mapped 
to generalized SpMV operations~\cite{GraphMat}.

\subsection{Performance Speedup}
\begin{figure}[t]
	\begin{center}
		\includegraphics[width=0.5\textwidth]{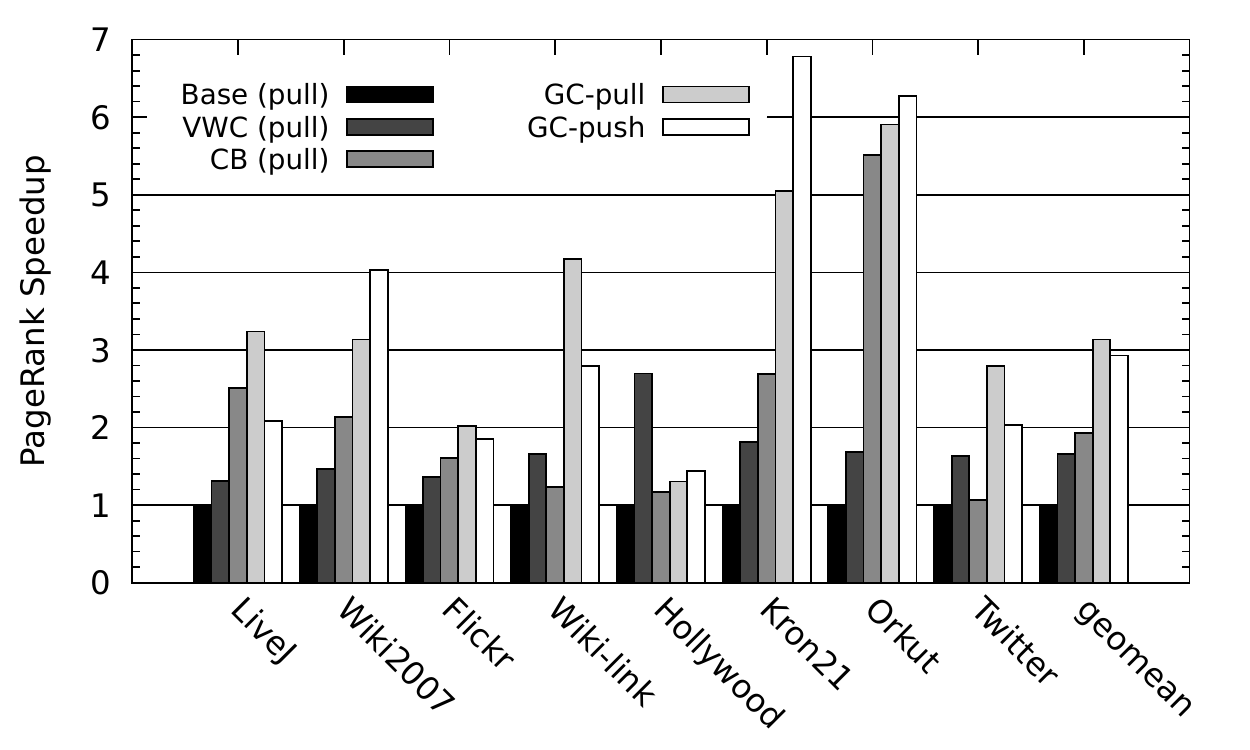}
		\vspace{-0.6cm}
		\caption{Performance of PR implementations, 
			all normalized to the baseline.}
		\label{fig:speedup-pr}
		\vspace{-0.5cm}
	\end{center}
\end{figure}

We first show PR performance improvement by comparing GraphCage
with hand-optimized implementations. The baseline (Base) is 
a straightforward pull-style implementation with no optimizations.
We then apply Virtual-Warp Centric (VWC)~\cite{Hong} method 
on the baseline, which is the second implementation. CB is the 
naive implementation of conventional cache blocking. We 
only implement cache blocking for PR. We also implement 
both PR-pull and PR-push in GraphCage. We run all the PR
implementations until they converge with the same condition.
\cref{fig:speedup-pr} illustrates the performance speedups
normalized to the baseline. On average, GraphCage significantly 
outperforms other implementations. VWC can consistently 
improve performance with an averaged speedup of 66\% because 
of coalesced memory accesses. But it has little effect on 
temporal locality. CB works well for relatively small graphs 
with poor locality. For large graphs, such as \texttt{Wiki-link}
and \texttt{Twitter}, it brings very limited benefit. For
\texttt{Hollywood}, however, neither CB and GraphCage are 
able to get too much speedup, since this graph already has 
a layout with good locality. But GraphCage can substantially
improve performance for large graphs, due to a better overhead
control of our proposed TOCAB scheme. Since our cache 
blocking approach works well with load balancing,
GraphCage consistently outperforms VWC (2$\times$ faster on 
average) except \texttt{Hollywood} which has a good layout 
that can benefit a lot from VWC, but not from locality 
optimizations. GC-pull and GC-push get similar average 
speedups, but have some difference for specific graphs. 
This is because GC-push has more overhead on atomic operations 
but its load balancing scheme is better than GC-pull.

\begin{figure}[t]
	\begin{center}
		\includegraphics[width=0.5\textwidth]{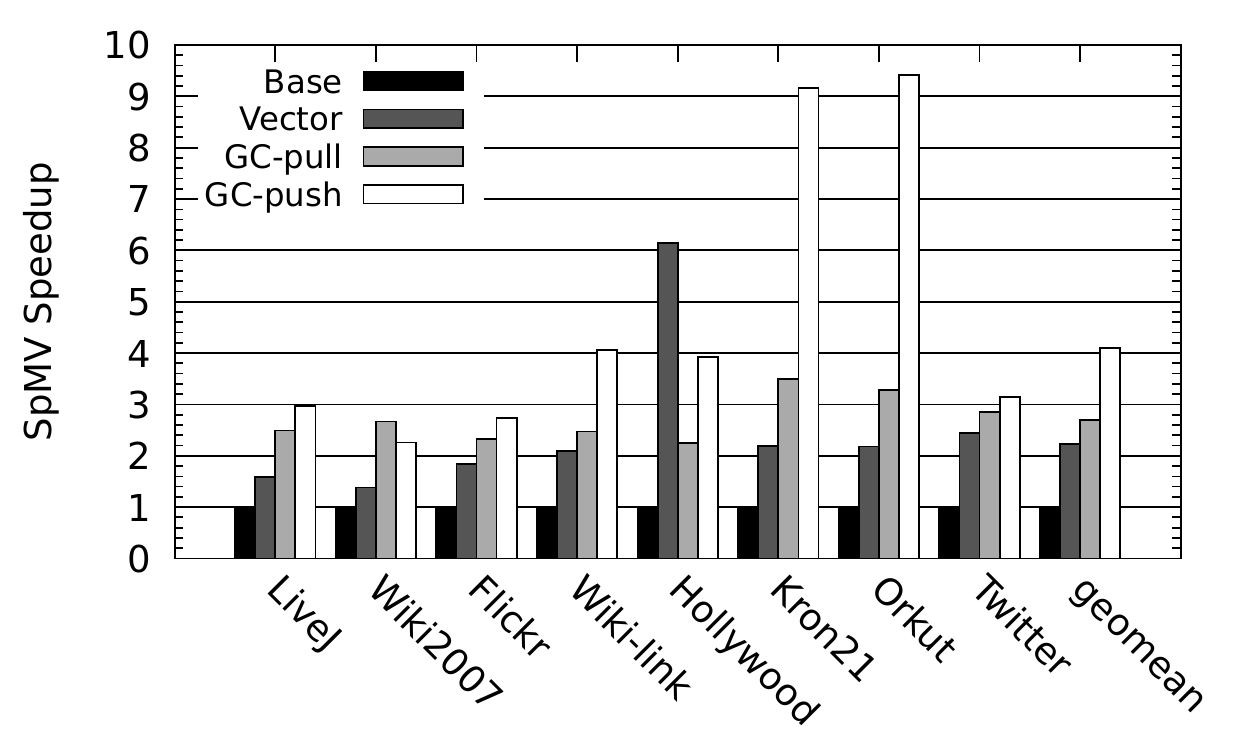}
		\vspace{-0.6cm}
		\caption{Performance of SpMV implementations, 
			all normalized to the baseline.}
		\label{fig:speedup-spmv}
		\vspace{-0.3cm}
	\end{center}
\end{figure}

We also evaluate the performance of SpMV and BC. \cref{fig:speedup-spmv} 
shows the speedups of SpMV implementations over the baseline. We can
still observe consistent performance improvement for VWC and GraphCage.
Compared to the baseline, VWC achieves more speedup (2.3$\times$ on 
average) for SpMV than PR because not only the column indices but 
also the matrix values are coalesced. The matrix values have no reuse
opportunity, and could pollute the last level cache by replacing useful
data. Therefore, GC-pull is less effective for SpMV compared to VWC.
Since the effect of load balancing for SpMV is more significant than
that for PR, we can observe better performance for GC-push than GC-pull,
since fine-grained load balancing is applied to GC-push, but not GC-pull.
On average, GC-push achieves 82\% performance improvement over VWC.

\begin{figure}[t]
	\begin{center}
		\includegraphics[width=0.5\textwidth]{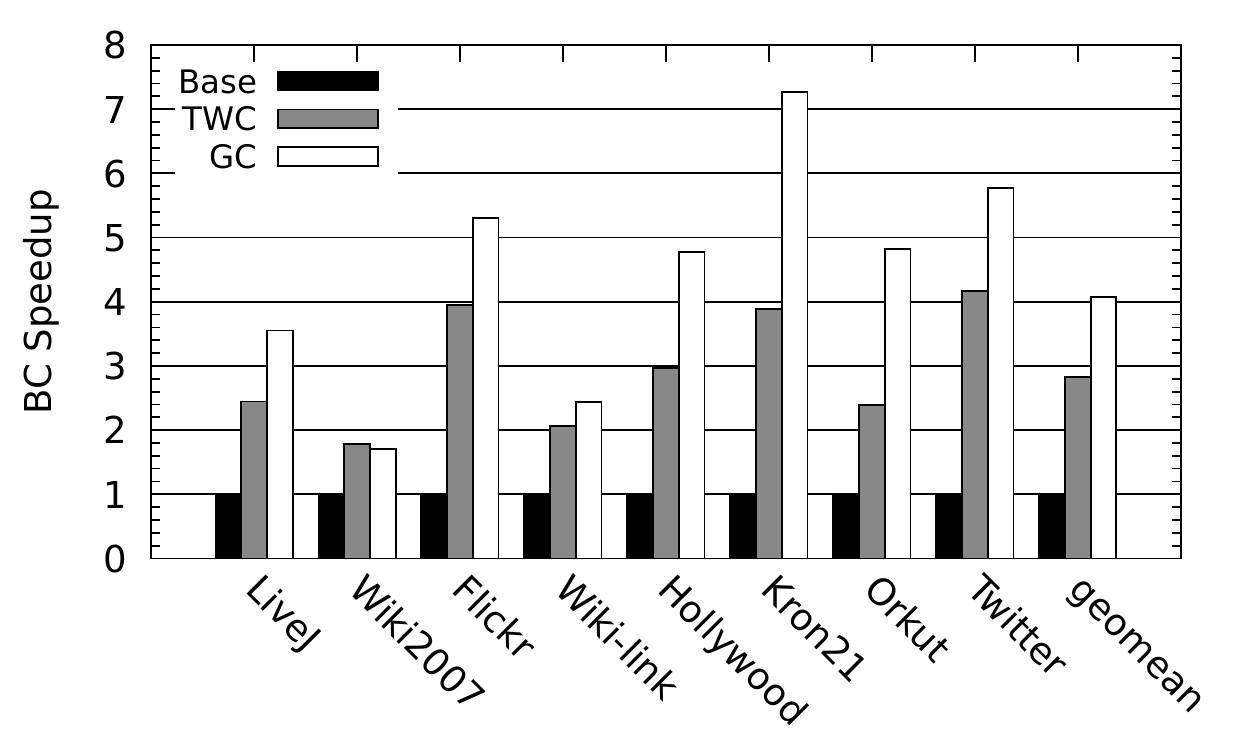}
		\vspace{-0.6cm}
		\caption{Performance of BC implementations, 
			all normalized to the baseline.}
		\label{fig:speedup-bc}
		\vspace{-0.3cm}
	\end{center}
\end{figure}

\cref{fig:speedup-bc} illustrates BC performance normalized to the 
baseline. Since we only applied TOCAB to the pull kernels, we have only 
one implementation for BC in GraphCage. The baseline is still a naive
implementation, but we apply TWC in the second implementation since TWC
works better than VWC for BC. We can still observe consistent improvement
for TWC and GraphCage with averaged speedup of 2.8$\times$ and 4.1$\times$. 
A special issue here is that for large-diameter graphs, \texttt{Wiki2007}
and \texttt{Wiki-link}, GraphCage has limited improvement over TWC. This
is because there are lots of iterations in these graphs have only one
vertex in the frontier, For these iterations, GraphCage does not bring
any benefit. Time spent on these iterations makes the improvement on 
other iterations less significant. However, we can still observe an 
average speedup of 46\% for GraphCage compared to TWC.

\subsection{Cache and Memory Efficiency}
\begin{figure}[t]
	\begin{center}
		\includegraphics[width=0.5\textwidth]{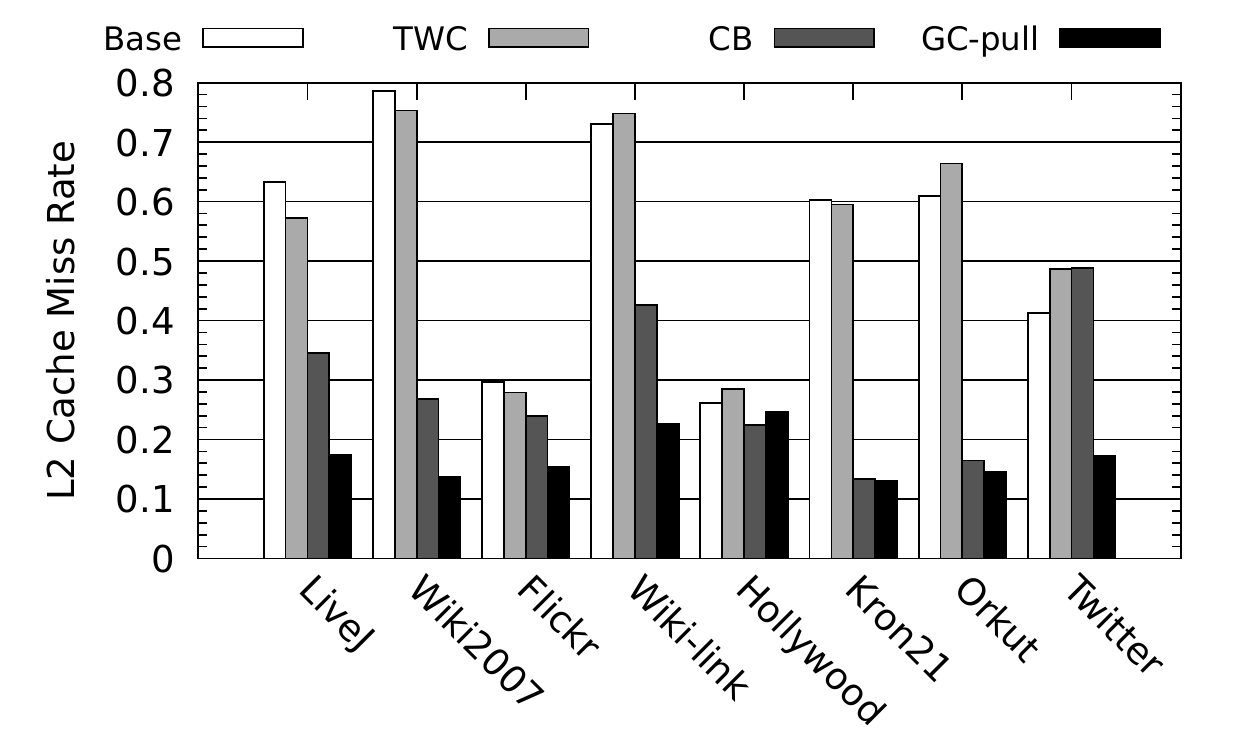}
		\vspace{-0.6cm}
		\caption{L2 cache miss rates of PR implementations.}
		\label{fig:l2-miss-rate}
		\vspace{-0.3cm}
	\end{center}
\end{figure}

To relate the total performance improvement to the cache 
behavior, we conduct experiments on cache miss rates and memory
efficiency. \cref{fig:l2-miss-rate} shows the L2 cache miss rates
for all the implementations in \cref{fig:speedup-pr}. It is quite
clear that Base and TWC suffer tremendous cache misses with up 
to about 80\% miss rate. In this case, the L2 cache is definitely
poorly used and it could become a bottleneck in the memory hierarchy.
Cache blocking techniques, however, can substantially reduce L2
cache miss rate. CB performs very well for \texttt{Kron21} and 
\texttt{Orkut} which have plenty of data reuses. But for other
graphs with poor locality, GraphCage can further reduce the miss 
rates to blow 20\%. For \texttt{Twitter}, since it is large enough
to have more than 80 subgraphs, CB suffer a huge overhead on 
repeated accesses and it fails to improve the cache behavior. 

\begin{figure}[t]
	\begin{center}
		\includegraphics[width=0.5\textwidth]{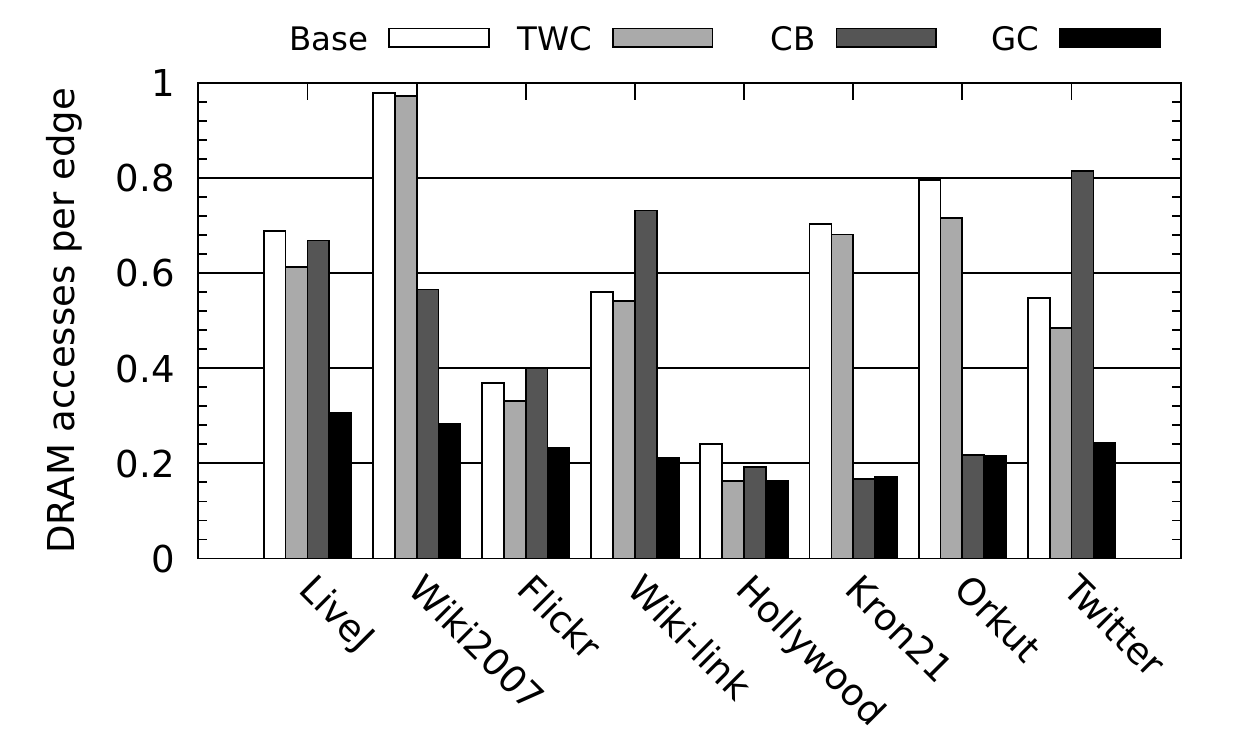}
		\vspace{-0.6cm}
		\caption{DRAM access efficiency (number of DRAM transactions per edge) 
		of different PR implementations.}
		\label{fig:dram}
		\vspace{-0.3cm}
	\end{center}
\end{figure}

We further look into the DRAM access efficiency by calculating the
number of DRAM accesses per edge. This ratio is one of the GAIL metrics 
to measure memory efficiency~\cite{GAIL}. As shown in \cref{fig:dram},
GraphCage significantly improves memory efficiency since accesses to
the vertex values are kept in the L2 cache, reducing a huge amount of
DRAM accesses. Although CB works well for \texttt{Kron21} and 
\texttt{Orkut}, it actually increase the total number of DRAM accesses
because of the repeated accesses to \texttt{sums}. This observation
explains the reason why CB reduces L2 cache miss rate for 
\texttt{Wiki-link} but the performance improvement is very limited.
The overhead of repeated accesses in CB is more significant for
larger graphs, leading to even more DRAM accesses than the baseline.
GraphCage, however, consistently achieves high memory efficiency.

\subsection{Sensitivity to Subgraph Size}
We vary the subgraph size to determine its impact on GraphCage's performance.
When the subgraphs are small enough to keep the vertex values reside in the 
cache, the memory efficiency is high, since most of the vertex value accesses
hit in the cache instead of going to the DRAM. However, if the subgraphs are
too small, there will be numerous subgraphs, which brings more overhead on 
merging the partial results of the subgraphs. Thus the performance will not 
be good. On the other hand, when the subgraph size becomes too large, the 
vertex values data no longer fit in the cache, and the memory efficiency
drops increasingly. Therefore, the subgraph size is a trade-off between 
locality benefits and the overhead of partitioning. For our platform, we
select a subgraph size of 256 vertices, since it makes a good trade-off
and achieves best performance.

\begin{figure}[t]
	\begin{center}
		\includegraphics[width=0.5\textwidth]{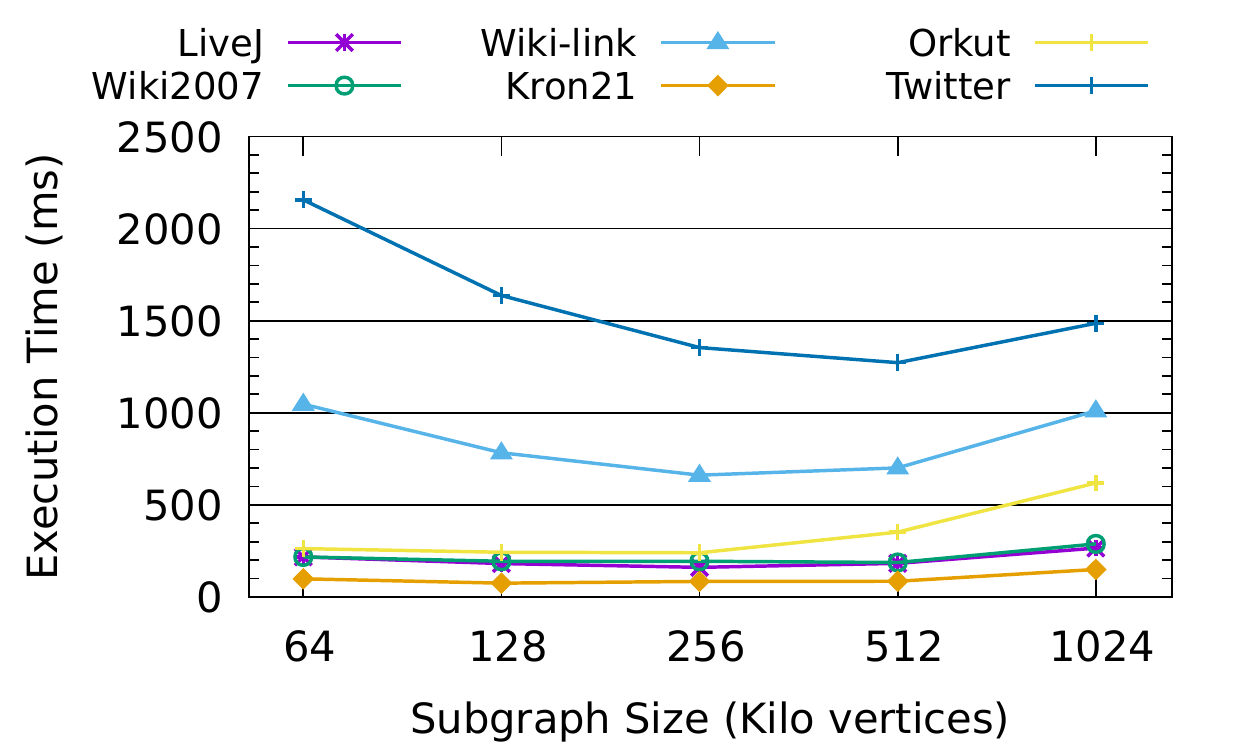}
		\vspace{-0.3cm}
		\caption{Impact of subgraph size on execution time of PR in 
			GraphCage. We select 256 vertices per subgraph due to 
			consistently good performance.}
		\label{fig:blksize}
		\vspace{-0.6cm}
	\end{center}
\end{figure}

\subsection{Comparing with Other Frameworks}
We compare GraphCage with the state-of-the-art GPU graph processing
framework Gunrock~\cite{Gunrock}, and another graph partitioning 
framework on GPUs, CuSha~\cite{CuSha}. The averaged execution time 
of one PageRank iteration is listed in Table~\ref{table:compare}.
We use the runtime of one iteration instead of runtime of convergence
because CuSha use a different convergence condition. We can observe
that GraphCage substantially outperforms Gunrock for all the graphs
with poor locality, especially those graphs with large average 
degrees which have great potential to enjoy locality optimizations.
On average, GC-pull and GC-push achieve 4.0$\times$ and 3.6$\times$ 
speedup over Gunrock for our tested graphs, respectively.
Since there is no cache blocking technique applied in Gunrock to 
improve cache performance, this dramatic improvement is expected.
For CuSha, however, it partitions graphs into Shards to fit into
the shared memory, which also improves performance for \texttt{LiveJ},
\texttt{Kron21} and \texttt{Orkut}. It is extremely inefficient 
for the two wiki graphs which have large diameters. This issue could
be caused by the difference of convergence condition. However,
GraphCage is still faster or as fast as CuSha for other graphs,
because of the overhead of relatively small partitions in CuSha.
Table~\ref{table:partition} list the number of partitions in
GraphCage and CuSha. Last but not least, CuSha fails to run 
Twitter on GTX 1080Ti because it requires more global memory than 
the GPU provides. This is also very important as global memory 
in GPUs is scarce resource and GraphCage can provide more 
stable performance with less global memory requirement than CuSha.
We also compare GraphCage with other available implementations.
For example, we compare SpMV with cuSPARSE~\cite{cuSPARSE} and 
nvGRAPH~\cite{nvGRAPH}. GraphCage also significantly outperforms 
these commercial libraries.

\begin{table}[]
	\small
	\centering
	\begin{tabular}{lllll}
		\Xhline{2\arrayrulewidth}
		& GC-pull & GC-push & Gunrock & CuSha  \\
		\hline
		LiveJ     & 5.33   & 8.26   & 29.04  & 9.30  \\
		Wiki2007  & 6.89   & 5.82   & 21.29  & 42.10 \\
		Flickr    & 2.62   & 3.06   & 5.89   & 12.88 \\
		Wiki-link & 27.53  & 41.19  & 124.40 & 438.72\\
		Hollywood & 8.36   & 7.60   & 8.69   & 9.55  \\
		Kron21    & 12.02  & 8.95   & 100.67 & 8.07  \\
		Orkut     & 11.40  & 10.73  & 116.65 & 10.75 \\
		Twitter   & 46.69  & 67.76  & 190.48 & out of memory\\
		\Xhline{2\arrayrulewidth}
	\end{tabular}
	\vspace{0.3cm}
	\caption{Averaged execution time (ms) of one PageRank iteration. 
		GraphCage significantly outperforms prior work.}
\vspace{-0.37cm}
\label{table:compare}
\end{table}

\begin{table}[t]
	\small
	\centering
	\begin{tabular}{c c c}
		\Xhline{2\arrayrulewidth}
		& \texttt{Subgraphs} & \texttt{Shards}\\
		\hline
		\textbf{LiveJ}     & 19 & 789 \\
		\textbf{Wiki2007}  & 14 & 1162\\
		\textbf{Flickr}    & 9  & 750 \\
		\textbf{Wiki-link} & 47 & 1978\\
		\textbf{Hollywood} & 5  & 743 \\
		\textbf{Kron21}    & 8  & 1366\\
		\textbf{Orkut}     & 12 & 1952\\
		\textbf{Twitter}   & 82 & 3467\\
		\Xhline{2\arrayrulewidth}
	\end{tabular}
	\vspace{0.3cm}
	\caption{Comparing the number of subgraphs in GraphCage and the 
		number of Shards in CuSha. Since GraphCage tiles the data 
		for the L2 cache, it has much fewer partitions.}
	\vspace{-0.5cm}
	\label{table:partition}
\end{table}

\section{Related Work}\label{sect:relate}
There are plenty of previous work on dynamic cache blocking targeting 
push-style graph processing on CPUs. Beamer~\emph{et~al.}~\cite{Beamer} 
proposed Propagation Blocking (PB)~\cite{PropBlocking} for PageRank. 
They split the computation into two phases, binning and accumulation. 
In the binning phase, intermediate data is distributed into different 
buffers (bins) whose sizes can fit in the cache. And then buffers are 
processed sequentially in the accumulation phase to merge the results. 
Similar approach is used to improve the performance of SpMV kernel in 
graph analytics applications~\cite{SpMV}. \texttt{milk} is a language 
extension on OpenMP~\cite{milk}. It automatically collects indirect 
memory accesses at runtime in order to improve locality. Comparing to 
these dynamic schemes on CPUs, our design use static cache blocking 
to reduce runtime overhead as much as possible.

Zhang~\emph{et~al.} propose CSR segmenting~\cite{Zhang}, an efficient 1D 
static cache blocking technique for multicore CPUs. CSR segmenting partitions
a graph into segments, each of which can fit in the cache. They also propose 
a merge phase to accumulate the partial results. Gluon~\cite{Gluon} is a 
graph processing system targeting heterogeneous distributed-memory platforms.
Gluon statically partitions the graph into subgraphs to fit into distributed
memories instead of caches. Their partitioning strategies for share-memory 
CPUs and distributed memory systems inspired our work. GraphCage is also a 
static cache blocking scheme, but we optimize it for the GPU architecture and 
integrate it with previously proposed optimization techniques for GPU graph processing.

GraphChi~\cite{GraphChi} is a edge centric graph processing system which
divides edges into shards to improve data locality on CPUs. 
CuSha~\cite{CuSha} extends this concept to GPU graph processing. They 
optimize it for the GPU execution model and use a COO-like graph format
called Concatenated Windows (CW) to store edges consecutively. This is 
particularly beneficial for GPU processing as memory accesses are fully 
coalesced. However, CW representation requires roughly 2.5x the size of 
CSR, which is not acceptable for GPUs whose global memory capacity is 
already very limited. By contrast, GraphCage requires much less memory 
space as we directly extends the data representation from CSR. Another 
issue is that Shards are constructed to fit into shared memory instead 
of LLC, which limits the subgraph size and brings more overhead.

There are also cache blocking schemes designed for other architectures. 
Graphphi~\cite{Graphphi} designed cache blocking for Intel Xeon Phi processors.
They proposed a hierarchical partitioning scheme to divide a graph into groups, 
stripes and tiles respectively. They use COO format instead of CSR to store 
the graph, and optimize the processing for Xeon Phi's MIMD-SIMD execution model.
Graphicionado~\cite{Graphicionado} is a customized hardware accelerator
for graph processing. It provides a large eDRAM on-chip as a scratchpad
memory to the programmers to explicitly arrange data accesses. The authors 
propose a graph partitioning scheme to keep data in the scratchpad, 
substantially reduce the memory access latency. However their scheme only 
discusses push-style graph processing, and is specifically optimized for 
their accelerator architecture. In comparison, GraphCage is optimized for 
GPUs and handles applications in both push and pull directions.

Beamer~\emph{et~al.}~\cite{Beamer} analyzed data locality of graph algorithms 
on CPUs. They found that there are actually plenty of data reuses in graph
algorithms, but existing hardware is not able to capture them, which leads
to poor cache performance. Xu~\emph{et~al.}~\cite{Xu} evaluated cache 
performance of graph processing on GPUs. They found that GPUs also suffer
this problem. Their observation motivates our work. Since hardware is 
unaware of the memory access pattern of graph algorithms, we started to
look into the software layer and rearrange the memory access order.
\section{Conclusion}\label{sect:concl}
Data locality is a key performance issue for graph processing.
In this paper, we point out that there is a chance to improve
locality as well as performance of GPU graph processing by 
leveraging the last level cache. We first propose a throughput 
oriented cache blocking (TOCAB) approach for the GPU architecture 
which improves memory access efficiency compared to naively 
applying conventional cache blocking scheme. Then we coordinate 
our cache blocking scheme with state-of-the-art load balancing 
scheme with awareness of the sparsity of subgraphs. Finally we 
apply our scheme to traversal based graph algorithm by considering 
both the benefit and overhead of cache blocking. These optimization 
techniques are integrated into our graph processing framework, i.e. 
GraphCage, to support various graph applications. Experimental 
results show that GraphCage can significantly improve performance 
over previous best-performing implementations and state-of-the-art 
graph processing frameworks such as Gunrock and CuSha, with 
less memory space requirement than CuSha. With the continuous 
increasing of GPU last level cache size, our proposed approach
will be more effective and important for graph processing on 
GPUs in the future.

\bibliographystyle{acmart}
\bibliography{references}
\end{document}